# Semiparametric integrative interaction analysis for non-small-cell lung cancer

Yang Li[1,2,3], Fan Wang[2,3], Rong Li[2,3] and Yifan Sun[1,2]

## Abstract
In genomic analysis, it is significant though challenging to identify markers associated with cancer outcomes or phenotypes. Based on the biological mechanisms of cancers and the characteristics of datasets, we propose a novel integrative interaction approach under a semiparametric model, in which genetic and environmental factors are included as the parametric and nonparametric components, respectively. The goal of this approach is to identify the genetic factors and gene–gene interactions associated with cancer outcomes, while estimating the nonlinear effects of environmental factors. The proposed approach is based on the threshold gradient-directed regularisation technique. Simulation studies indicate that the proposed approach outperforms alternative methods at identifying the main effects and interactions, and has favourable estimation and prediction accuracy. We analysed non-small-cell lung carcinoma datasets from the Cancer Genome Atlas, and the results demonstrate that the proposed approach can identify markers with important implications and that it performs favourably in terms of prediction accuracy, identification stability, and computation cost.

## 1 Introduction

Lung cancer is the second most common cancer in the world, accounting for over 1.6 million deaths annually.[1] Approximately 85% of lung cancer is classified as non-small-cell lung cancer (NSCLC).[2] The Cancer Genome Atlas (TCGA), a collaborative effort organised by the National Cancer Institute (NCI), published high-quality profiling data on multiple cancer types, including NSCLC. The resulting rich data provide a major opportunity to uncover the underlying genetic factors of cancers. From the viewpoint of statistical analysis, however, it is a challenging task to identify the markers associated with outcomes and phenotypic variances of NSCLC.

### 1.1 Multiple datasets

The first challenge stems from the data. NSCLC data are classified into two subtypes: lung adenocarcinoma (LUAD) and lung squamous cell carcinoma (LUSC). These are the two most prevalent subtypes, which account for the majority of lung cancer diagnoses worldwide. In clinical studies of NSCLC, LUAD and LUSC are commonly discussed together to find similarities and differences in their molecular organisation, activity, and gene expression profile.[3,4] Integrating the two datasets of LUAD and LUSC can help us gain a comprehensive understanding of the genomic basis of NSCLC. Integrative approaches to multiple datasets are broadly categorised into three categories.[5] Early integration combines all datasets directly into one and analyses the combined

[1]Center for Applied Statistics, Renmin University of China, Beijing, China
[2]School of Statistics, Renmin University of China, Beijing, China
[3]Statistical Consulting Center, Renmin University of China, Beijing, China

**Corresponding author:**
Yifan Sun, Center for Applied Statistics, Renmin University of China, 59 Zhongguancun Street, Haidian District, Beijing, China.
Email: sunyifan1984@163.com

dataset as a whole. However, because one gene can have different effects on different cancer subtypes, the early integration approach neglects the intrinsic heterogeneity of the two subtypes, leading to low reliability of the analysis results. With late integration, i.e. meta-analysis, each dataset is analysed independently, and the results are then combined across datasets. However, both the LUAD and LUSC datasets have 'small $n$, large $p$' characteristics ($p \sim 18,000$ and $n \sim 200$), leading to unsatisfactory results for each individual dataset, and hence the overall meta-analysis.[6] The third category is intermediate integration, which is often called 'integrative analysis' in biostatistics. Integrative analysis aims to preserve the structure of multiple datasets and only merges them during the modelling process. This has been shown to outperform other multi-dataset integration methods.[7,8] In this paper, we adopt the integrative analysis approach to handle multiple datasets.

Within an integrative analysis framework, multiple datasets can be described using the homogeneity structure or the heterogeneity structure.[7] Under the homogeneity structure, the same set of genetic factors is identified across multiple datasets. The heterogeneity structure differs from the homogeneity structure by allowing multiple datasets to have different sets of important genetic factors. The data analysed in this study are composed of two datasets corresponding to the two subtypes of NSCLC. The common biological mechanism underlying diverse subtypes of the same cancer is such that it is reasonable to expect that each gene exerts either no effect or significant but varying effects on the different subtypes. To this end, we performed integrative analysis under the homogeneity structure. That is, we identified the same set of genetic factors for multiple subtypes of the same cancer, while allowing for different magnitudes of effects.

## 1.2 Gene–gene interactions

Possible gene–gene interactions pose new challenges to data analysis. Accumulating evidence suggests that gene–gene interactions contribute to explaining and predicting disease outcomes or phenotypes.[9,10] The introduction of gene–gene interactions into the statistical model significantly increases the number of covariates, and, hence, aggravates the high dimensionality issue. Consider a dataset with $n$ samples and $p$ genetic measurements. In interaction analysis, the total number of unknown parameters is $\frac{p(p+1)}{2}$, which often exceeds $n$ even for a moderate $p$. Moreover, it has been a widely recognised practice that a statistical model with interactions should meet the strong hierarchical constraint. That is, if an interaction is identified, then the two main effects involved must also be identified.[11] Under the strong hierarchical constraint, if $k < p$, genes are expected to be relevant to the cancer outcome, and there are at most $\frac{k(k-1)}{2}$ interactions associated with the outcome. Hence, there is a selection problem. It is unfeasible to apply traditional variable-selection methods directly, as they may violate this hierarchical structure. For a single dataset, delicate variable-selection methods that ensure a hierarchical structure have been proposed.[11] In this study, we conducted integrative interaction analysis while satisfying the hierarchical structure of selection results in multiple datasets simultaneously.

## 1.3 Environmental factors

Like genetic factors, many environmental factors have non-negligible effects on cancer outcomes. For example, smoking is by far the leading cause of lung cancer,[12] and age is found to be associated with the development and progression of lung cancer.[13] Unlike genetic measurements, however, environmental factors often display non-linear relationships with cancer outcomes. Figure 1 shows scatterplots and fitted curves of smoking and age against the percentage reference values for the pre-bronchodilator forced expiratory volume in one second (FEV1) in LUAD and LUSC. It is clear that, except for the curve of age in LUAD, the curves have significantly nonlinear trends. For this reason, we propose a semiparametric approach to analysing NSCLC data with genetic factors as the parametric parts and environmental factors as the nonparametric parts. Nonparametric functions are approximated using the B-spline technique. Variable selection is only conducted for the parametric parts.

In this paper, motivated by the NSCLC data in TCGA, we propose an integrative interaction analysis approach under a semiparametric model. The proposed approach identifies genetic factors and gene–gene interactions associated with disease outcomes while estimating the nonlinear effects of environmental factors. Compared to standard integrative analysis approaches that have been developed to analyse cancer data,[7,8] the proposed approach considers gene–gene interactions and some environmental risk factors. Building on existing approaches,[9,10] our proposal jointly models multiple datasets, and helps to reveal common mechanisms as well as dataset-specific cancer genomic characteristics. Li et al.[14] analysed NSCLC data with $p = 100$ genes in a semi-parametric model by using a penalisation method. They demonstrated the effectiveness of the penalisation method for that data. This paper aims to analyse a 'larger' dataset ($p = 300$). Indeed, genetic datasets are often large in scale, with which a penalisation method will encounter convergence problems. As shown in Figure 2, the estimated coefficients of PYGB, a randomly selected gene, cannot converge to a fixed point. This result motivated us to develop a new approach with good convergence ability beyond the penalised method. The proposed approach is based on the threshold gradient directed regularisation (TGDR) technique, which is popular in high-dimensional analysis.[15] Despite it being similar to some penalisation methods such as fused lasso, TGDR is a completely different method of conducting variable selection and estimation. It offers widespread applications, high efficiency, and robust performance, and it has been extensively discussed in literature.[16–18] In particular, TGDR offers good convergence with NSCLC data (Figure 2).

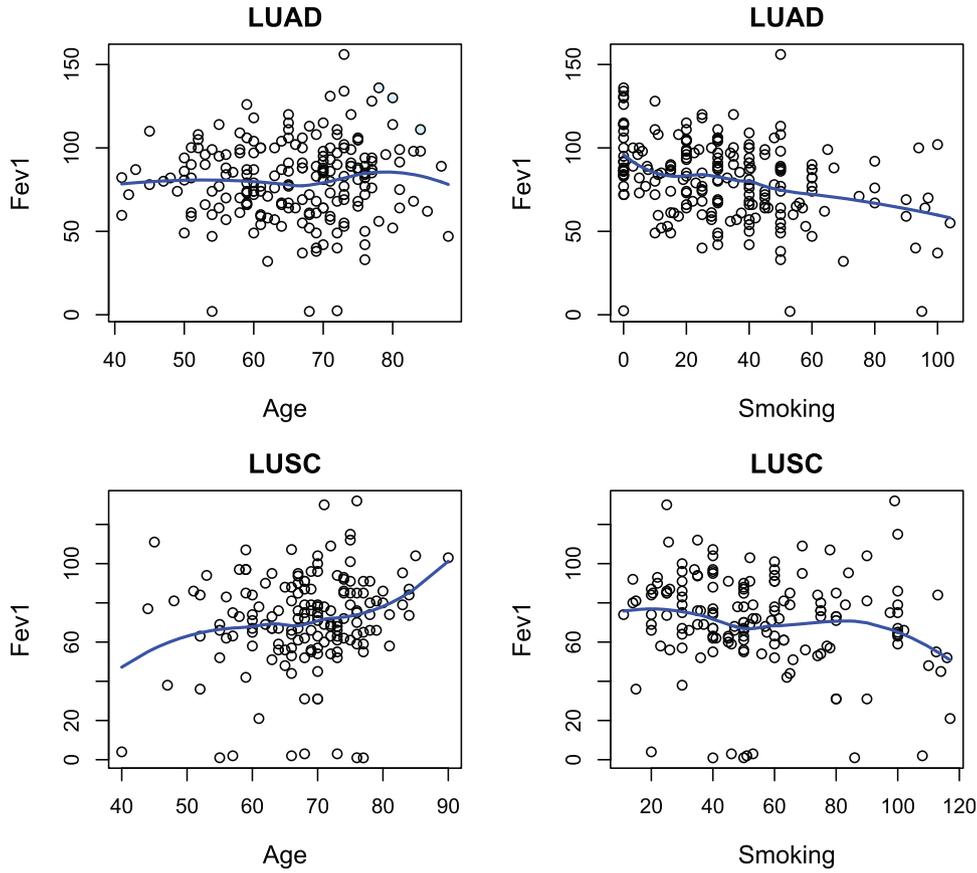

**Figure 1.** Age and smoking versus FEV1. The blue line denotes a smoothed locally weighted scatterplot.

From the perspective of methodology, we developed a novel TGDR for semiparametric integrative interaction analysis, as an extension to the work by Li et al.[19] Moreover, data on other cancers in TCGA have characteristics similar to those of NSCLC: multiple datasets, high dimensionality, a small sample size, gene–gene interactions, and environmental factors. Consequently, the proposed approach can be seen as a general method of analysing cancer data.

The remainder of this article is organised as follows. Section 2 introduces the model, method, and algorithm. In Section 3, we describe our evaluation of the performance of the proposed method with extensive simulations. We present our analysis of NSCLC data with two types of response variables in Section 4. Finally, Section 5 offers our conclusions. Additional technical details and numerical results are provided in the supplementary Appendix.

## 2 Methodology

### 2.1 Model

Suppose there are $M$ independent datasets, each with $n_m$ ($m = 1, 2, \ldots, M$) i.i.d. observations. Each dataset corresponds to one subtype of cancer. Let $\boldsymbol{Y}^{(m)} = \left(y_1^{(m)}, y_2^{(m)}, \ldots, y_{n_m}^{(m)}\right)^\top$ denote the response variable, and $\boldsymbol{X}^{(m)} = (\boldsymbol{X}_1^{(m)}, \boldsymbol{X}_2^{(m)}, \ldots, \boldsymbol{X}_p^{(m)})$ with $\boldsymbol{X}_j^{(m)} = (X_{1j}^{(m)}, X_{2j}^{(m)}, \ldots, X_{n_m j}^{(m)})^\top$ denote the measurements of $p$ genes. The interaction of genes $j$ and $k$ is therefore defined as $\boldsymbol{X}_j^{(m)} \boldsymbol{X}_k^{(m)} = \left(X_{1j}^{(m)} X_{1k}^{(m)}, X_{2j}^{(m)} X_{2k}^{(m)}, \ldots, X_{n_m j}^{(m)} X_{n_m k}^{(m)}\right)^\top$. For simplicity, we assume that the same set of genes is measured in all $M$ datasets. Let us represent by $\boldsymbol{E}^{(m)} = \left(\boldsymbol{E}_1^{(m)}, \boldsymbol{E}_2^{(m)}, \ldots, \boldsymbol{E}_q^{(m)}\right)$ the $q$-dimensional environmental risk factors, where $\boldsymbol{E}_l^{(m)} = \left(E_{1l}^{(m)}, E_{2l}^{(m)}, \ldots, E_{n_m l}^{(m)}\right)^\top$. The semiparametric integrative interaction analysis model is formulated as follows

$$\boldsymbol{Y}^{(m)} \sim \Phi\left(\sum_{j=1}^{p} \beta_j^{(m)} \boldsymbol{X}_j^{(m)} + \sum_{j<k} \gamma_{jk}^{(m)} \boldsymbol{X}_j^{(m)} \boldsymbol{X}_k^{(m)} + \sum_{l=1}^{q} \theta_l\left(\boldsymbol{E}_l^{(m)}\right)\right) \quad (1)$$

where $\beta_j^{(m)}$ is the unknown coefficient of gene $j$ in dataset $m$, and $\gamma_{jk}^{(m)}$ is the corresponding coefficient of the interaction term $\boldsymbol{X}_j^{(m)} \boldsymbol{X}_k^{(m)}$. The link function $\Phi$ is assumed to have a known form, and $\theta_l(\cdot)$ is an unknown smoothing function. Here, the data have been normalised such that there is no intercept in model (1).

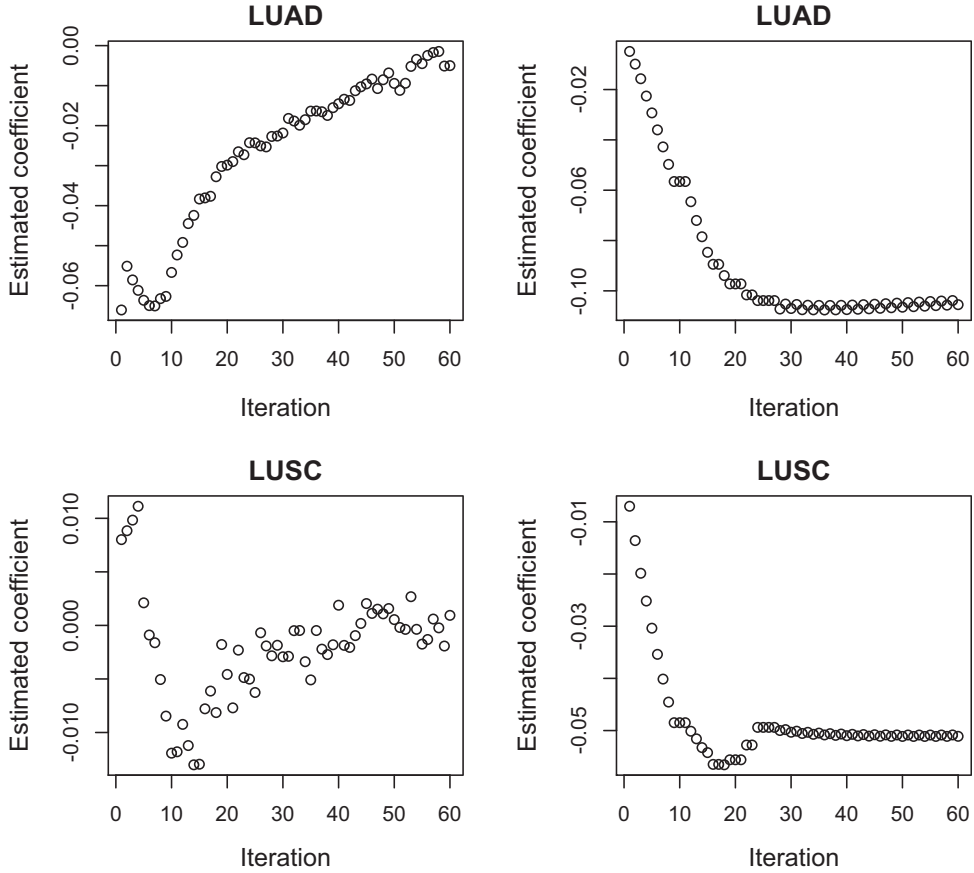

**Figure 2.** Analysis of NSCLC data (with survival time as the response variable): estimated coefficient of gene PYGB over iteration. Left: penalisation method. Right: proposed method.

We adopt the spline technique to approximate the unknown function $\theta_l(\cdot)$ ($l = 1, \ldots, q$) with cubic B-spline as the basis function. Assume that $\boldsymbol{E}_l^{(m)}$ ($l = 1, 2, \ldots, q$) are continuous variables, and that the function $\theta_l^{(m)}$ satisfies $\int_0^1 \theta_l^{(m)}(E)\,\mathrm{d}E = 0$. Under the smoothness assumptions stated by the Curry–Schorberg theorem, each smoothing function $\theta_l^{(m)}(\cdot)$ can be approximated by

$$\theta_l^{(m)}(\cdot) \approx \sum_{d=1}^{D} \eta_{ld}^{(m)} B_d(\cdot) = \boldsymbol{B}(\cdot) \boldsymbol{\eta}_l^{(m)}$$

where $\boldsymbol{B} = (B_1, B_2, \ldots, B_D)$ is an equally spaced B-spline basis, and $\boldsymbol{\eta}_l^{(m)} = \left(\eta_{l1}^{(m)}, \eta_{l2}^{(m)}, \ldots, \eta_{lD}^{(m)}\right)$ is a length-$D$ vector of coefficients. Then model (1) can be approximated by

$$Y^{(m)} \sim \Phi\left(\sum_{j=1}^{p} \beta_j^{(m)} X_j^{(m)} + \sum_{j<k} \gamma_{jk}^{(m)} X_j^{(m)} X_k^{(m)} + \sum_{l=1}^{q} B\left(E_l^{(m)}\right) \boldsymbol{\eta}_l^{(m)}\right)$$

To simplify the notations, let $\boldsymbol{\beta}^{(m)} = \left(\beta_1^{(m)}, \beta_2^{(m)}, \ldots, \beta_p^{(m)}\right)^\top$, $\boldsymbol{\gamma}^{(m)} = \left(\gamma_{12}^{(m)}, \gamma_{13}^{(m)}, \ldots, \gamma_{p-1,p}^{(m)}\right)^\top$ and $\boldsymbol{Z}^{(m)} = \left(\boldsymbol{\eta}_1^{(m)}, \ldots, \boldsymbol{\eta}_q^{(m)}\right)^\top$. We denote by $L^{(m)}$ the loss function (or other lack-of-fitness measures) of dataset $m$. When $\Phi$ is a linear regression model, for example, the least squares loss function takes the following form

$$L^{(m)}\left(\boldsymbol{\beta}^{(m)}, \boldsymbol{\gamma}^{(m)}, \boldsymbol{Z}^{(m)}\right) = \left\|Y^{(m)} - \sum_{j=1}^{p} \beta_j^{(m)} X_j^{(m)} - \sum_{j<k} \gamma_{jk}^{(m)} X_j^{(m)} X_k^{(m)} - \sum_{l=1}^{q} B\left(E_l^{(m)}\right) \boldsymbol{\eta}_l^{(m)}\right\|_2^2$$

In this study, we also consider right-censored survival data under the accelerated failure time (AFT) model. Details on the model settings and loss functions are provided in supplementary Appendix A.

## 2.2 Algorithm

We adopt the TGDR method to select important covariates, estimate unknown coefficients and predict outcomes. TGDR, a regularisation method, was first proposed in a linear regression model with high-dimensional covariates by Friedman and Popescu,[15] and then generalised to other types of models, such as the Cox model,[17] generalised linear model,[16,20] and single-index model.[18] The basic idea of this method is to first define a set of candidate models as a path through the space of joint parameter values, and then choose a point on this path to be the final model by minimising an appropriate objective function. Compared to other types of regularisation methods, TGDR has many desirable advantages, including generality, robustness, high speed, and satisfactory prediction performance.[15] Li et al.[19] proposed TGDR for integrative interaction analysis. The model they considered, however, is a parametric one. As such, TGDR is unable to analyse NSCLC data. Thus, we developed a new TGDR method.

Under the assumption of a semiparametric model, TGDR for integrative interaction analysis consists of the following iterative steps:

1. Initialisation. We set $t=0$. We denote by $\boldsymbol{\beta}^{(m)}(t)$, $\boldsymbol{\gamma}^{(m)}(t)$, and $\boldsymbol{Z}^{(m)}(t)$ the estimates of $\boldsymbol{\beta}^{(m)}$, $\boldsymbol{\gamma}^{(m)}$, and $\boldsymbol{Z}^{(m)}$ in the $t$th iteration, respectively. We then initialise $\boldsymbol{\beta}^{(m)}(t) = 0$, $\boldsymbol{\gamma}^{(m)}(t) = 0$, and $\boldsymbol{Z}^{(m)}(t) = 0$, respectively.
2. Computing gradients. We update $t = t + 1$. For $m = 1, \ldots, M$, we compute the negative gradients

$$f_j^{(m)} = -\frac{\partial L^{(m)}}{\partial \beta_j^{(m)}}\Big|_{\beta_j^{(m)} = \beta_j^{(m)}(t-1)}$$

$$g_{jk}^{(m)} = -\frac{\partial L^{(m)}}{\partial \gamma_{jk}^{(m)}}\Big|_{\gamma_{jk}^{(m)} = \gamma_{jk}^{(m)}(t-1)}$$

$$\boldsymbol{h}^{(m)} = -\frac{\partial L^{(m)}}{\partial \boldsymbol{Z}^{(m)}}\Big|_{\boldsymbol{Z}^{(m)} = \boldsymbol{Z}^{(m)}(t-1)}$$

3. Computing the indicators $\tilde{\boldsymbol{f}} = (\tilde{f}_1, \ldots, \tilde{f}_p)^\top$ with $\tilde{f}_j = \mathbb{I}\left(\sum_{m=1}^{M} \left|f_j^{(m)}\right| > \tau \times \max_u \sum_{m=1}^{M} \left|f_u^{(m)}\right|\right)$, and $\tilde{\boldsymbol{g}} = (\tilde{g}_{12}, \ldots, \tilde{g}_{p-1,p})^\top$ with $\tilde{g}_{jk} = \mathbb{I}\left(\sum_{m=1}^{M} \left|g_{jk}^{(m)}\right| > \tau \times \max_{u,v} \sum_{m=1}^{M} \left|g_{uv}^{(m)}\right|\right)$. Here $\tilde{g}_{jk}$ and $\tilde{f}_j$ ($\tilde{f}_k$) follow the strong hierarchy restriction. That is, if $\tilde{g}_{jk} = 1$, then $\tilde{f}_j = 1$, and $\tilde{f}_k = 1$.

4. Updating.

$$\begin{aligned}\boldsymbol{\beta}^{(m)}(t) &= \boldsymbol{\beta}^{(m)}(t-1) + \Delta \times \tilde{\boldsymbol{f}} \; o \; \boldsymbol{f}^{(m)}\\ \boldsymbol{\gamma}^{(m)}(t) &= \boldsymbol{\gamma}^{(m)}(t-1) + \Delta \times \tilde{\boldsymbol{g}} \; o \; \boldsymbol{g}^{(m)}\\ \boldsymbol{Z}^{(m)}(t) &= \boldsymbol{Z}^{(m)}(t-1) + \Delta \times \boldsymbol{h}^{(m)}\end{aligned}$$

where the step size $\Delta$ is set to 0.01.

5. Iterating. Repeat Steps 2–4 $T$ times.

In Step 4, $o$ denotes the Hadamard product. The threshold $\tau$ controls the degree of regularisation, and, in this study, it is fixed to 0.9. Compared to standard TGDR algorithms, the proposed algorithm has three unique characteristics. First, in Step 3, the indicators are determined by considering all $M$ datasets simultaneously. Second, the algorithm ensures that the strong hierarchy constraint is satisfied. That is, if an interaction is selected, then its corresponding two main effects are also selected. Finally, the update of nonparametric spline parameters is added as a new part of the algorithm. Parametric and nonparametric components are treated in different ways. Owing to the high dimensionality of parametric components, we select important parametric components and estimate their coefficients. However, the dimensions of nonparametric components are kept low. This is true with many datasets containing real data, as for example, with the NSCLC data analysed in this study. Consequently, we here leave nonparametric components unselected. The number of iterations was selected using five-fold cross-validation.

## 3 Numerical study

### 3.1 Simulation settings

We simulated $M=3$ independent datasets with sample sizes $n_1 = 180$, $n_2 = 170$, and $n_3 = 150$, respectively (total sample size $n = 500$). There were $q = 2$ environmental factors, and $p$ genetic factors with $p = 50$ and 100, respectively. The genetic factors had marginally normal distributions with mean 0 and variance 1. In each dataset, there were 10 important genetic factors and 10 important gene–gene interactions. We considered eight cases for the coefficients of parametric components: $(\underbrace{\beta_1,\ldots,\beta_p}_{\text{main effects}}, \underbrace{\gamma_{12},\ldots,\gamma_{p-1,p}}_{\text{interactions}})$. The values of nonzero coefficients in Cases I–V were as follows

I
| | $\beta_1$ | $\beta_2$ | $\beta_3$ | $\beta_4$ | $\beta_5$ | $\beta_6$ | $\beta_7$ | $\beta_8$ | $\beta_9$ | $\beta_{10}$ | $\gamma_{12}$ | $\gamma_{13}$ | $\gamma_{14}$ | $\gamma_{23}$ | $\gamma_{24}$ | $\gamma_{34}$ | $\gamma_{56}$ | $\gamma_{57}$ | $\gamma_{67}$ | $\gamma_{89}$ |
|---|---|---|---|---|---|---|---|---|---|---|---|---|---|---|---|---|---|---|---|---|
| Data 1 | 2 | 2 | 2 | 2 | 2 | 1 | 1 | 1 | 1 | 1 | 2 | 1.5 | 1.5 | 1 | 1 | 1 | 0.5 | 0.5 | 0.5 | 0.5 |
| Data 2 | 1.5 | 1.5 | 1.5 | 1.5 | 1.5 | 1 | 1 | 1 | 1 | 1 | 2 | 1 | 1 | 0.5 | 0.5 | 0.5 | 1.5 | 1.5 | 1.5 | 1.5 |
| Data 3 | 1 | 1 | 1 | 1 | 1 | 0.5 | 0.5 | 0.5 | 0.5 | 0.5 | 1.5 | 1 | 1 | 0.5 | 0.5 | 0.5 | 2 | 2 | 2 | 2 |

II
| | $\beta_1$ | $\beta_2$ | $\beta_3$ | $\beta_4$ | $\beta_5$ | $\beta_6$ | $\beta_7$ | $\beta_8$ | $\beta_9$ | $\beta_{10}$ | $\gamma_{12}$ | $\gamma_{13}$ | $\gamma_{14}$ | $\gamma_{23}$ | $\gamma_{24}$ | $\gamma_{34}$ | $\gamma_{56}$ | $\gamma_{57}$ | $\gamma_{67}$ | $\gamma_{89}$ |
|---|---|---|---|---|---|---|---|---|---|---|---|---|---|---|---|---|---|---|---|---|
| Data 1 | 1 | 1 | 1 | 1 | 1 | 1 | 1 | 1 | 1 | 1 | 2 | 1.5 | 1.5 | 1 | 1 | 1 | 0.5 | 0.5 | 0.5 | 0.5 |
| Data 2 | 1.5 | 1.5 | 1.5 | 1.5 | 1.5 | 1 | 1 | 1 | 1 | 1 | 2 | 1 | 1 | 0.5 | 0.5 | 0.5 | 1.5 | 1.5 | 1.5 | 1.5 |
| Data 3 | 1.5 | 1.5 | 1.5 | 1.5 | 1.5 | 0.5 | 0.5 | 0.5 | 0.5 | 0.5 | 1.5 | 1 | 1 | 0.5 | 0.5 | 0.5 | 1 | 1 | 1 | 2 |

III
| | $\beta_1$ | $\beta_2$ | $\beta_3$ | $\beta_4$ | $\beta_5$ | $\beta_6$ | $\beta_7$ | $\beta_8$ | $\beta_9$ | $\beta_{10}$ | $\gamma_{12}$ | $\gamma_{13}$ | $\gamma_{14}$ | $\gamma_{23}$ | $\gamma_{24}$ | $\gamma_{34}$ | $\gamma_{56}$ | $\gamma_{57}$ | $\gamma_{67}$ | $\gamma_{89}$ |
|---|---|---|---|---|---|---|---|---|---|---|---|---|---|---|---|---|---|---|---|---|
| Data 1 | 2 | 2 | 2 | 2 | 2 | $-1$ | 1 | 1 | 1 | 1 | 2 | 1.5 | 1.5 | $-1$ | 1 | 1 | 0.5 | 0.5 | 0.5 | 0.5 |
| Data 2 | 1.5 | 1.5 | 1.5 | 1.5 | 1.5 | 1 | 1 | 1 | 1 | 1 | 2 | 1 | 1 | 0.5 | 0.5 | 0.5 | 1.5 | 1.5 | 1.5 | 1.5 |
| Data 3 | 1 | 1 | 1 | 1 | $-1$ | 0.5 | 0.5 | 0.5 | 0.5 | 0.5 | 1.5 | 1 | 1 | 0.5 | 0.5 | 0.5 | $-2$ | 2 | 2 | 2 |

$$\text{IV} \begin{pmatrix} & \beta_1 & \beta_2 & \beta_3 & \beta_4 & \beta_5 & \beta_6 & \beta_7 & \beta_8 & \beta_9 & \beta_{10} & \gamma_{12} & \gamma_{13} & \gamma_{14} & \gamma_{23} & \gamma_{24} & \gamma_{34} & \gamma_{56} & \gamma_{57} & \gamma_{67} & \gamma_{89} \\ \text{Data 1} & 2 & 2 & -2 & 2 & 2 & -1 & 1 & 1 & 1 & 1 & 2 & -1.5 & 1.5 & 1 & 1 & 1 & 0.5 & 0.5 & 0.5 & 0.5 \\ \text{Data 2} & -1.5 & -1 & 1 & -1.5 & 1.5 & 1 & 1 & 1 & 1 & 1 & 2 & 1 & 1 & 0.5 & 0.5 & 0.5 & 1.5 & 1.5 & 1.5 & 1.5 \\ \text{Data 3} & -0.5 & -1 & 1 & -0.5 & -1 & 0.5 & 0.5 & 0.5 & 0.5 & 0.5 & 1.5 & 0.5 & 1 & 0.5 & 0.5 & 0.5 & -2 & 2 & 2 & 2 \end{pmatrix}$$

$$\text{V} \begin{pmatrix} & \beta_1 & \beta_2 & \beta_3 & \beta_4 & \beta_5 & \beta_6 & \beta_7 & \beta_8 & \beta_9 & \beta_{10} & \gamma_{12} & \gamma_{13} & \gamma_{14} & \gamma_{15} & \gamma_{16} & \gamma_{17} & \gamma_{18} & \gamma_{19} & \gamma_{1(10)} & \gamma_{1(11)} \\ \text{Data 1} & 2 & 2 & 2 & 2 & 2 & 1 & 1 & 1 & 1 & 1 & 2 & 1.5 & 1.5 & 1 & 1 & 1 & 0.5 & 0.5 & 0.5 & 0.5 \\ \text{Data 2} & 1.5 & 1.5 & 1.5 & 1.5 & 1.5 & 1 & 1 & 1 & 1 & 1 & 2 & 1 & 1 & 0.5 & 0.5 & 0.5 & 1.5 & 1.5 & 1.5 & 1.5 \\ \text{Data 3} & 1 & 1 & 1 & 1 & 1 & 0.5 & 0.5 & 0.5 & 0.5 & 0.5 & 1.5 & 1 & 1 & 0.5 & 0.5 & 0.5 & 2 & 2 & 2 & 2 \end{pmatrix}$$

The genetic factors were generated independently. The strong hierarchical constraint was satisfied in all cases, except in Case V. Compared to Case I, the nonzero coefficients in Case II had less variation across the three datasets, whereas those in Case III had more variation. Even coefficients of the same covariate had different signs across datasets. In Case IV, the sum of coefficients of the same covariate could be zero, suggesting that the overall effect of this covariate could be cancelled out. In Case VI, the genetic factors had an autoregressive correlation structure with the correlation coefficient of genes $j$ and $k$ as $0.5^{|j-k|}$. Other settings were the same as those in Case I. The only difference between Cases VII and I was that, in Case VII, the main effects and interactions were estimated independently without meeting the strong hierarchical constraint.

For the environmental factors, we considered two scenarios: they were nonlinearly (a) and linearly (b) associated with the response variables. In (a), the environmental factors were included in the model as nonparametric parts, and the true model was semiparametric. Specifically, Cases I–VII had the same nonparametric functions in the three datasets

$$\eta_1(E) = sin(4\pi E), \quad \eta_2(E) = 10(e^{-3.25E} + 4e^{-6.5E} + 3e^{-9.75E})$$

In Case VIII, the nonparametric functions differed in the three datasets

$$\begin{aligned} \eta_1^{(1)}(E) &= \sin(4\pi E), & \eta_2^{(1)}(E) &= 10(e^{-3.25E} + 4e^{-6.5E} + 3e^{-9.75E}) \\ \eta_1^{(2)}(E) &= \sin(4\pi E), & \eta_2^{(2)}(E) &= 3(2E-1)^2 \\ \eta_1^{(3)}(E) &= 10(e^{-3.25E} + 4e^{-6.5E} + 3e^{-9.75E}), & \eta_2^{(3)}(E) &= 3(2E-1)^2 \end{aligned}$$

and the parametric settings were the same as those in Case I. For scenario (b), the environmental measurements were treated as parametric parts, and the true model was a parametric one. The nonparametric functions were

$$\eta_1(E) = E, \quad \eta_2(E) = E$$

for all three datasets in cases I–VII, and

$$\begin{aligned} \eta_1^{(1)}(E) &= \eta_2^{(1)}(E) = E + 1 \\ \eta_1^{(2)}(E) &= \eta_2^{(2)}(E) = 0.5E + 1 \\ \eta_1^{(3)}(E) &= \eta_2^{(3)}(E) = E + 1.5 \end{aligned}$$

in Case VIII. For all settings, the environmental factors were generated independently from the uniform distribution $\mathcal{U}(0,1)$.

To better gauge the performance of the proposed method, we compared it to three alternatives. (1) Meta-analysis ('Meta') involves conducting semiparametric TGDR on each dataset separately. (2) Pool-analysis ('Pool') combines all datasets directly into one and applies semiparametric TGDR. (3) Parametric-analysis ('Parametric') involves TGDR under a parametric model,[19] which treats all predictors as parametric ones. We measured the performance of effect identification, including identification of the main effects and interactions, according to the true positive rate (TPR) and false positive rate (FPR), of which the definitions were the same as those given in the literature. The prediction and estimation performance was also evaluated. Specifically, the prediction performance was measured by the prediction error (PE), which is defined as $\frac{1}{n}\sum_{m=1}^{M}\|Y^{(m)} - \hat{Y}^{(m)}\|_2^2$. The estimation performance was measured by the mean squared error (MSE) of the coefficients of parametric components, defined as $\frac{1}{n}\sum_{m=1}^{M}\|\boldsymbol{\beta}^{(m)} - \hat{\boldsymbol{\beta}}^{(m)}\|_2^2$ (main effects) and $\frac{1}{n}\sum_{m=1}^{M}\|\boldsymbol{\gamma}^{(m)} - \hat{\boldsymbol{\gamma}}^{(m)}\|_2^2$ (interactions), and by the root-mean-square error (RMSE) for the nonparametric spline parameters, defined as $\sqrt{\frac{1}{n}\sum_{m=1}^{M}\|\mathbf{Z}^{(m)} - \hat{\mathbf{Z}}^{(m)}\|_2^2}$.

## 3.2 Results

A summary of the statistics was computed based on 100 independent replicates. The results of the linear regression model are shown in Tables 1 and 2, corresponding to the true model being semiparametric and parametric, respectively. The measurement errors were generated from $N(0, 1)$. When the data were generated from the semiparametric model, the proposed method was observed to have superior performance at identifying interactions—for example, in Case III in Table 1. When identifying interactions, the proposed method had (TPR, FPR) = (0.970, 0.000), which was superior to the alternatives: (0.419, 0.001) for Meta, (0.770, 0.002) for 'Pool', and (0.895, 0.000) for 'Parametric'. For a more comprehensive evaluation of the identification performance, we considered a sequence of tunings, and provide partial ROC curves for each method (Figure B.1 in supplementary Appendix B). The results show that the proposed method indeed has advantages when identifying interactions. For identifications of main effects, the proposed method and 'Parametric' performed similarly, outperforming the other two alternatives. For example, in Case IV, the proposed method and 'Parametric' had (TPR, FPR) = (0.986, 0.048) and (0.975, 0.033), respectively, compared to (0.703, 0.088) for 'Meta' and (0.852, 0.525) for 'Pool'. When structured correlations between the covariates were introduced (Case VI), the FPRs of the main effects increased significantly with all four methods. The proposed method still performed well when identifying the main effects and interactions, even when the data did not meet the strong hierarchical constraint (Case V), implying the robustness of the method. Overall, 'Meta' had the worst performance owing to small sample sizes. As expected, 'Pool' deteriorated in performance when there were large differences in the values of coefficients across datasets (Case IV, for example). In terms of estimation and prediction, the proposed method outperformed the alternatives. In all cases considered here, the proposed method had the lowest estimation errors and prediction errors. From Table 2, it follows that the 'Parametric' method had the best performance. This result was not unexpected, because the data were generated from a parametric model. Nevertheless, the proposed method had comparable performance in this scenario. As such, it is a 'safe' choice in practice. Similar observations were also made under other settings, as presented in supplementary Appendix B.

We also plotted the nonparametric fitting curves and the 95% point-wise confidence intervals using the bootstrap method (see supplementary Appendix B). In Case I, fitting curves produced by the proposed method were close to the true ones, and the confidence intervals contained the true curves completely, regardless of whether the true model was adopted. Other cases had similar results. Owing to space constraints, however, we do not describe them here.

We also conducted simulations under the AFT model. The censoring rates were around 10%. The covariates were generated in the same way as in Case III, and the environmental factors were modelled as nonparametric components. The simulation results are provided in Table B.3 (supplementary Appendix B). The proposed method was again observed to have favourable performance in terms of identification, estimation, and prediction. In addition, we also conducted a simulation to examine the robustness of the proposed method to the signal-to-noise ratio. Specifically, two scenarios were considered: (S1) with measurement errors generated from $N(0, 4)$, and (S2) with coefficients divided by 2 and measurement errors still generated from $N(0, 1)$. Table B.3 presents the results of Case III under the linear regression model. Overall, the proposed method had favourable performance for $p = 50$, although its identification performance suffered to some extent for $p = 100$, though it still outperformed the alternative methods in terms of estimation and prediction.

## 4 Analysis of NSCLC data

In this section, we apply the proposed method to NSCLC data. As described in Section 1, in TCGA, there are two NSCLC datasets: LUAD and LUSC. On the one hand, because the two subtypes belong to NSCLC, it is expected

Table 1. Simulation results: $p = 50$, and the data are generated from the semiparametric model.

| Cases | | Main | | | Interaction | | | RMSE | PE |
|---|---|---|---|---|---|---|---|---|---|
| | | TPR | FPR | MSE | TPR | FPR | MSE | | |
| I | Proposed | 0.989 (0.035) | 0.046 (0.071) | 0.044 (0.023) | 0.964 (0.067) | 0.000 (0.000) | 0.005 (0.003) | 0.657 (0.093) | 2.272 (0.836) |
| | Meta | 0.692 (0.139) | 0.080 (0.060) | 0.525 (0.155) | 0.405 (0.131) | 0.001 (0.001) | 0.024 (0.006) | 1.181 (0.138) | 15.547 (4.653) |
| | Pool | 1.000 (0.000) | 0.095 (0.153) | 0.105 (0.025) | 0.992 (0.031) | 0.002 (0.002) | 0.008 (0.001) | 0.467 (0.079) | 4.890 (0.683) |
| | Parametric | 0.974 (0.048) | 0.033 (0.049) | 0.082 (0.033) | 0.891 (0.102) | 0.000 (0.000) | 0.009 (0.003) | 1.475 (0.062) | 5.743 (1.147) |
| II | Proposed | 0.998 (0.014) | 0.067 (0.085) | 0.030 (0.013) | 0.977 (0.051) | 0.000 (0.000) | 0.004 (0.001) | 0.619 (0.084) | 1.977 (0.458) |
| | Meta | 0.732 (0.108) | 0.092 (0.076) | 0.336 (0.108) | 0.453 (0.116) | 0.001 (0.001) | 0.016 (0.004) | 1.027 (0.122) | 10.145 (3.209) |
| | Pool | 1.000 (0.000) | 0.255 (0.220) | 0.054 (0.011) | 1.000 (0.000) | 0.001 (0.001) | 0.005 (0.001) | 0.375 (0.071) | 3.065 (0.325) |
| | Parametric | 0.991 (0.029) | 0.038 (0.051) | 0.061 (0.022) | 0.913 (0.102) | 0.000 (0.000) | 0.007 (0.002) | 1.469 (0.065) | 5.332 (0.930) |
| III | Proposed | 0.991 (0.029) | 0.040 (0.052) | 0.039 (0.019) | 0.970 (0.052) | 0.000 (0.000) | 0.005 (0.002) | 0.633 (0.090) | 2.258 (0.735) |
| | Meta | 0.689 (0.142) | 0.068 (0.067) | 0.527 (0.150) | 0.419 (0.115) | 0.001 (0.001) | 0.024 (0.005) | 1.156 (0.137) | 15.854 (4.575) |
| | Pool | 0.974 (0.044) | 0.059 (0.090) | 0.266 (0.035) | 0.770 (0.069) | 0.002 (0.002) | 0.016 (0.002) | 0.624 (0.116) | 9.902 (1.142) |
| | Parametric | 0.976 (0.049) | 0.022 (0.030) | 0.081 (0.034) | 0.895 (0.110) | 0.000 (0.000) | 0.009 (0.003) | 1.464 (0.066) | 5.916 (1.368) |
| IV | Proposed | 0.986 (0.035) | 0.048 (0.079) | 0.054 (0.033) | 0.944 (0.083) | 0.000 (0.000) | 0.006 (0.003) | 0.686 (0.107) | 2.718 (1.033) |
| | Meta | 0.703 (0.153) | 0.088 (0.078) | 0.480 (0.159) | 0.415 (0.121) | 0.001 (0.001) | 0.023 (0.006) | 1.163 (0.139) | 14.911 (4.696) |
| | Pool | 0.852 (0.098) | 0.102 (0.150) | 0.620 (0.030) | 0.525 (0.114) | 0.000 (0.001) | 0.024 (0.003) | 0.832 (0.144) | 18.436 (2.111) |
| | Parametric | 0.975 (0.050) | 0.033 (0.064) | 0.095 (0.049) | 0.863 (0.119) | 0.000 (0.000) | 0.010 (0.003) | 1.469 (0.059) | 6.369 (1.358) |
| V | Proposed | 0.987 (0.039) | 0.062 (0.071) | 0.050 (0.037) | 0.929 (0.087) | 0.000 (0.000) | 0.007 (0.003) | 0.689 (0.115) | 2.766 (1.211) |
| | Meta | 0.621 (0.166) | 0.081 (0.075) | 0.571 (0.161) | 0.367 (0.128) | 0.001 (0.001) | 0.027 (0.006) | 1.213 (0.127) | 17.501 (5.101) |
| | Pool | 1.000 (0.000) | 0.148 (0.182) | 0.106 (0.030) | 0.992 (0.031) | 0.002 (0.003) | 0.008 (0.002) | 0.458 (0.087) | 4.904 (0.826) |
| | Parametric | 0.979 (0.043) | 0.044 (0.056) | 0.089 (0.044) | 0.862 (0.108) | 0.000 (0.000) | 0.010 (0.004) | 1.471 (0.071) | 6.188 (1.311) |
| VI | Proposed | 0.986 (0.038) | 0.253 (0.120) | 0.059 (0.031) | 0.993 (0.033) | 0.000 (0.000) | 0.003 (0.002) | 0.698 (0.130) | 2.472 (1.031) |
| | Meta | 0.980 (0.032) | 0.390 (0.103) | 0.142 (0.065) | 0.896 (0.079) | 0.002 (0.002) | 0.006 (0.003) | 0.913 (0.195) | 3.951 (2.690) |
| | Pool | 1.000 (0.000) | 0.325 (0.204) | 0.110 (0.022) | 0.991 (0.029) | 0.002 (0.005) | 0.008 (0.001) | 0.600 (0.107) | 8.660 (1.456) |
| | Parametric | 0.985 (0.041) | 0.254 (0.139) | 0.068 (0.034) | 0.991 (0.032) | 0.000 (0.000) | 0.003 (0.002) | 1.536 (0.066) | 4.380 (1.223) |
| VII | Proposed | 0.994 (0.024) | 0.023 (0.035) | 0.048 (0.027) | 0.976 (0.051) | 0.000 (0.001) | 0.005 (0.002) | 0.633 (0.095) | 2.239 (0.761) |
| | Meta | 0.581 (0.162) | 0.068 (0.065) | 0.584 (0.149) | 0.405 (0.124) | 0.001 (0.001) | 0.025 (0.006) | 1.221 (0.138) | 16.033 (4.440) |
| | Pool | 0.996 (0.024) | 0.048 (0.099) | 0.128 (0.048) | 0.997 (0.017) | 0.006 (0.007) | 0.008 (0.001) | 0.472 (0.086) | 5.047 (0.951) |
| | Parametric | 0.985 (0.039) | 0.021 (0.037) | 0.097 (0.048) | 0.926 (0.081) | 0.001 (0.001) | 0.008 (0.003) | 1.477 (0.066) | 5.722 (1.220) |
| VIII | Proposed | 0.984 (0.042) | 0.047 (0.067) | 0.049 (0.039) | 0.962 (0.064) | 0.000 (0.000) | 0.006 (0.003) | 0.652 (0.122) | 2.473 (1.110) |
| | Meta | 0.722 (0.119) | 0.082 (0.065) | 0.508 (0.147) | 0.418 (0.116) | 0.001 (0.001) | 0.023 (0.006) | 1.141 (0.133) | 14.853 (4.385) |
| | Pool | 0.998 (0.014) | 0.084 (0.133) | 0.107 (0.030) | 0.994 (0.024) | 0.001 (0.002) | 0.009 (0.002) | 0.839 (0.055) | 5.556 (0.737) |
| | Parametric | 0.978 (0.046) | 0.040 (0.053) | 0.084 (0.038) | 0.876 (0.110) | 0.000 (0.000) | 0.009 (0.003) | 1.522 (0.078) | 5.808 (1.329) |

Note: In each cell, we show the mean (SD).

**Table 2.** Simulation results: $p = 50$, and the data are generated from the parametric model.

| Cases | | Main | | | Interaction | | | | |
|---|---|---|---|---|---|---|---|---|---|
| | | TPR | FPR | MSE | TPR | FPR | MSE | RMSE | PE |
| I | Proposed | 0.992 (0.027) | 0.064 (0.085) | 0.034 (0.019) | 0.971 (0.052) | 0.000 (0.000) | 0.005 (0.002) | 0.395 (0.078) | 2.251 (0.741) |
| | Meta | 0.753 (0.147) | 0.116 (0.079) | 0.459 (0.169) | 0.462 (0.123) | 0.001 (0.001) | 0.022 (0.006) | 0.858 (0.175) | 14.042 (4.689) |
| | Pool | 1.000 (0.000) | 0.097 (0.141) | 0.100 (0.025) | 0.993 (0.026) | 0.001 (0.002) | 0.008 (0.001) | 0.297 (0.075) | 5.025 (0.701) |
| | Parametric | 0.990 (0.030) | 0.067 (0.079) | 0.032 (0.020) | 0.985 (0.039) | 0.000 (0.000) | 0.004 (0.002) | 0.186 (0.054) | 2.101 (0.680) |
| II | Proposed | 0.998 (0.014) | 0.060 (0.058) | 0.025 (0.012) | 0.983 (0.040) | 0.000 (0.000) | 0.004 (0.001) | 0.366 (0.061) | 1.970 (0.464) |
| | Meta | 0.786 (0.126) | 0.099 (0.080) | 0.287 (0.110) | 0.503 (0.124) | 0.001 (0.001) | 0.015 (0.004) | 0.760 (0.157) | 8.985 (3.081) |
| | Pool | 1.000 (0.000) | 0.207 (0.197) | 0.052 (0.011) | 1.000 (0.000) | 0.001 (0.001) | 0.005 (0.001) | 0.236 (0.061) | 3.060 (0.330) |
| | Parametric | 1.000 (0.000) | 0.065 (0.070) | 0.022 (0.009) | 0.989 (0.031) | 0.000 (0.000) | 0.004 (0.001) | 0.165 (0.051) | 1.916 (0.332) |
| III | Proposed | 0.994 (0.028) | 0.048 (0.073) | 0.035 (0.021) | 0.971 (0.050) | 0.000 (0.000) | 0.005 (0.002) | 0.400 (0.069) | 2.348 (0.819) |
| | Meta | 0.784 (0.126) | 0.109 (0.081) | 0.428 (0.161) | 0.492 (0.118) | 0.001 (0.001) | 0.021 (0.005) | 0.812 (0.165) | 13.081 (4.150) |
| | Pool | 0.980 (0.040) | 0.058 (0.105) | 0.264 (0.043) | 0.779 (0.062) | 0.002 (0.002) | 0.016 (0.002) | 0.409 (0.117) | 10.097 (1.203) |
| | Parametric | 0.996 (0.020) | 0.053 (0.062) | 0.030 (0.017) | 0.978 (0.044) | 0.000 (0.000) | 0.004 (0.002) | 0.185 (0.055) | 2.242 (0.732) |
| IV | Proposed | 0.990 (0.030) | 0.061 (0.085) | 0.040 (0.029) | 0.972 (0.057) | 0.000 (0.000) | 0.005 (0.003) | 0.402 (0.075) | 2.599 (1.142) |
| | Meta | 0.773 (0.132) | 0.122 (0.085) | 0.408 (0.147) | 0.478 (0.113) | 0.001 (0.001) | 0.021 (0.005) | 0.823 (0.164) | 13.382 (4.689) |
| | Pool | 0.860 (0.102) | 0.125 (0.142) | 0.615 (0.030) | 0.516 (0.122) | 0.000 (0.001) | 0.024 (0.004) | 0.546 (0.173) | 19.104 (2.239) |
| | Parametric | 0.995 (0.022) | 0.065 (0.087) | 0.032 (0.021) | 0.977 (0.051) | 0.000 (0.000) | 0.005 (0.002) | 0.184 (0.059) | 2.368 (0.913) |
| V | Proposed | 0.994 (0.028) | 0.056 (0.066) | 0.040 (0.021) | 0.940 (0.075) | 0.000 (0.000) | 0.007 (0.003) | 0.411 (0.077) | 2.694 (0.950) |
| | Meta | 0.735 (0.128) | 0.096 (0.059) | 0.467 (0.142) | 0.447 (0.113) | 0.001 (0.001) | 0.023 (0.005) | 0.858 (0.155) | 13.846 (3.955) |
| | Pool | 1.000 (0.000) | 0.112 (0.141) | 0.105 (0.029) | 0.985 (0.044) | 0.002 (0.002) | 0.009 (0.002) | 0.302 (0.078) | 5.095 (0.717) |
| | Parametric | 0.995 (0.022) | 0.053 (0.061) | 0.036 (0.018) | 0.934 (0.083) | 0.000 (0.000) | 0.007 (0.004) | 0.192 (0.064) | 2.863 (1.167) |
| VI | Proposed | 0.987 (0.034) | 0.254 (0.132) | 0.054 (0.022) | 0.997 (0.017) | 0.000 (0.000) | 0.003 (0.001) | 0.416 (0.096) | 2.468 (0.850) |
| | Meta | 0.990 (0.023) | 0.397 (0.116) | 0.121 (0.044) | 0.908 (0.066) | 0.002 (0.002) | 0.005 (0.002) | 0.536 (0.152) | 3.539 (1.600) |
| | Pool | 1.000 (0.000) | 0.354 (0.212) | 0.114 (0.024) | 0.993 (0.026) | 0.002 (0.004) | 0.008 (0.001) | 0.409 (0.115) | 8.534 (1.293) |
| | Parametric | 0.988 (0.033) | 0.269 (0.121) | 0.051 (0.022) | 0.999 (0.010) | 0.000 (0.000) | 0.002 (0.001) | 0.200 (0.055) | 2.245 (0.494) |
| VII | Proposed | 0.995 (0.022) | 0.030 (0.046) | 0.039 (0.020) | 0.970 (0.050) | 0.000 (0.000) | 0.005 (0.002) | 0.384 (0.074) | 2.401 (0.869) |
| | Meta | 0.671 (0.167) | 0.087 (0.070) | 0.501 (0.159) | 0.469 (0.128) | 0.002 (0.001) | 0.022 (0.006) | 0.852 (0.155) | 14.073 (4.601) |
| | Pool | 0.996 (0.032) | 0.100 (0.158) | 0.122 (0.058) | 0.998 (0.014) | 0.005 (0.006) | 0.008 (0.002) | 0.304 (0.086) | 5.053 (1.177) |
| | Parametric | 0.993 (0.026) | 0.039 (0.059) | 0.034 (0.020) | 0.982 (0.041) | 0.000 (0.000) | 0.004 (0.002) | 0.171 (0.052) | 2.129 (0.777) |
| VIII | Proposed | 0.990 (0.036) | 0.065 (0.078) | 0.034 (0.020) | 0.994 (0.024) | 0.000 (0.000) | 0.004 (0.002) | 0.456 (0.076) | 2.076 (0.557) |
| | Meta | 0.816 (0.129) | 0.140 (0.081) | 0.394 (0.187) | 0.530 (0.123) | 0.001 (0.001) | 0.019 (0.005) | 0.809 (0.160) | 11.769 (4.778) |
| | Pool | 1.000 (0.000) | 0.104 (0.165) | 0.097 (0.026) | 0.988 (0.033) | 0.001 (0.003) | 0.008 (0.001) | 0.503 (0.076) | 5.063 (0.743) |
| | Parametric | 0.988 (0.033) | 0.067 (0.081) | 0.034 (0.020) | 0.990 (0.030) | 0.000 (0.000) | 0.004 (0.002) | 0.288 (0.077) | 2.111 (0.512) |

Note: In each cell, we show the mean (SD).

Table 3. Analysis of NSCLC data (with FEV1 as the response variable) using the proposed method: estimated coefficients of main effects and interactions.

|  | LUAD | LUSC |  | LUAD | LUSC |
|---|---|---|---|---|---|
| EIF4A3 | −0.021 | −0.042 | METTL21C | 0.014 | 0.010 |
| DELE1 | 0.033 | 0.018 | IRS4 | −0.020 | −0.020 |
| TP63 | −0.024 | 0.012 | SLC27A2 | −0.017 | −0.054 |
| C5ORF38 | 0.038 | 0.042 | ZNF596 | 0.012 | 0.016 |
| CPSF3 | −0.011 | −0.019 | ARHGAP36 | −0.022 | −0.049 |
| DHX58 | 0.024 | 0.011 | CPSF3 × FBXO28 | −0.015 | −0.022 |
| ALPL | 0.065 | −0.035 | DHX58 × FBXO28 | −0.013 | 0.011 |
| IL6 | −0.011 | −0.028 | OSBPL7 × IRS4 | −0.021 | −0.014 |
| OSBPL7 | 0.016 | 0.012 | FBXO28 × METTL21C | 0.016 | 0.007 |
| FBXO28 | −0.021 | −0.042 | SLC27A2 × ARHGAP36 | 0.035 | 0.050 |

that they may share some biological mechanisms.[3] Consequently, we assume that the two subtypes have the same set of important genes. On the other hand, the differences between the two subtypes have already been acknowledged.[4] To this end, we allow each identified gene and interaction to have varying magnitudes of effects in the two subtypes.

In this analysis, we are interested in quantifying the influence of gene expressions and environmental risk factors on the forced expiratory volume in one second (FEV1), an important measure of lung function, and on the survival time of patients. A total of 18,277 gene expressions were collected for the patients in LUAD and LUSC. To tackle the high dimensionality and improve the reliability of the analysis, we conducted a marginal screening procedure and kept the first 300 genes with the smallest $p$-values. The total number of genetic factors was thus 45,150, including 300 main effects and 44,850 interactions. Motivated by existing studies and our preliminary analysis (Figure 1), age and smoking (measured by packs per year) were added as the environmental factors in the form of nonparametric functions.

### 4.1 FEV1 as the response variable

In the raw datasets, there were 517 patients in LUAD and 501 patients in LUSC. After removing missing values from FEV1, tumour stage, and environmental factors, a total of 378 patients (207 from LUAD and 171 from LUSC) were included. More details on sample selection are provided in Figure C.1 (supplementary Appendix C). The median FEV1 was 78 (range: 0.94–156), the median age at diagnosis was 68 (range: 40–90, year), and the median amount of smoking was 40 (range: 0–180, packs/year).

The proposed method identified 15 main effects and five interactions. The detailed estimation results are shown in Table 3. A literature review suggested that the findings were biologically meaningful. First, some identified genes are known as the oncogenes of lung cancers. For example, gene EIF4A3 is especially involved in the development of NSCLC.[21] Miki et al.[22] revealed that genetic variation in TP63 is significantly associated with lung adenocarcinoma susceptibility in Asian populations. Gene IL6 has been extensively studied. Research has found that IL6 blockage inhibits the promotion of disease, and a high level of IL6 has been observed in some lung cancer patients.[23] In the identified interactions, SLC27A2 expression was confirmed to have a reduction in CD166+ lung cancer stem cells of NSCLC samples.[24] The over-expression of IRS4 contributed to the promotion of tumours in lung cancers.[25]

Figure 3 displays estimates of nonparametric functions. The point-wise confidence intervals of curves were obtained using the bootstrap method over 50 random replicates drawn from the original datasets. All the functions diverged from a straight line, suggesting that the semiparametric model is more appropriate for describing the environmental effects on FEV1 in our data.

We also adopted the alternative methods described above to this data to generate different findings. Table 4 provides the summary of the comparison. Meta-analysis identified 19 (16) main effects, and 7 (3) interactions in LUAD (LUSC). Only two main effects were shared by the two subtypes. Pool-analysis identified 18 main effects and 5 interactions. With its particular property, the coefficients of the identified main effects and interactions were the same in the two subtypes. Parametric-analysis identified 13 main effects and 3 interactions, of which the magnitudes differed among the two subtypes. Less overlap between the proposed method and 'Parametric' in the

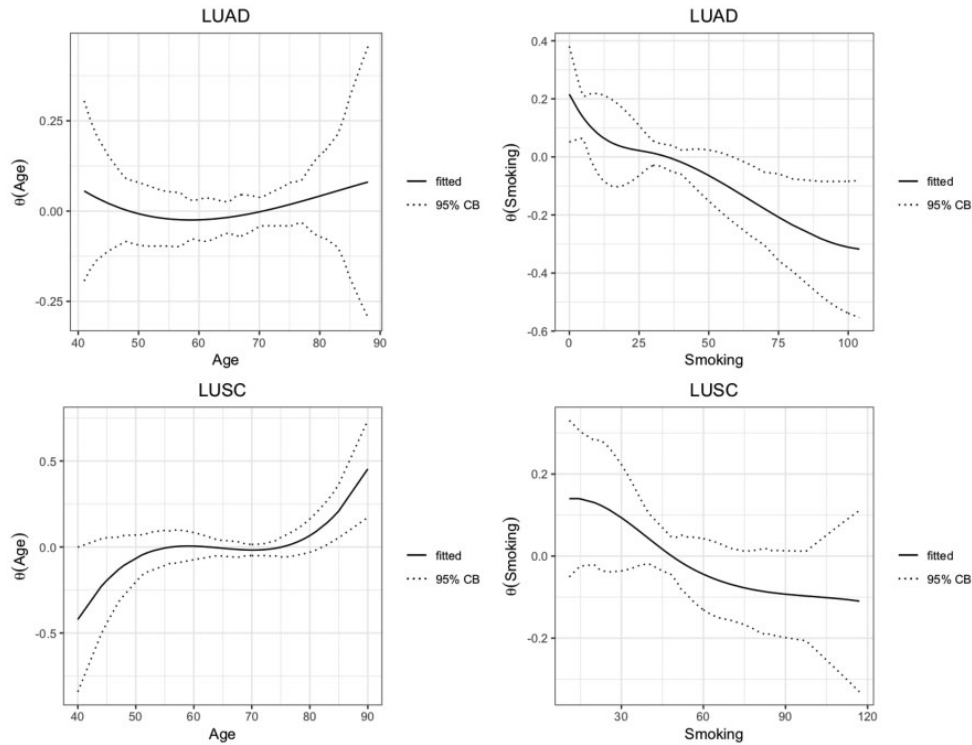

**Figure 3.** Analysis of NSCLC data (with FEV1 as the response variable) using the proposed method: estimated nonparametric functions of age and smoking with 95% confidence bands.

**Table 4.** Analysis of NSCLC data (with FEV1 as the response variable): number of overlapping main effects and interactions.

|  |  | Proposed | Meta | Pool | Parametric |
|---|---|---|---|---|---|
| Main effects | Proposed | 15/15 | 3/5 | 10/10 | 1/1 |
|  | Meta |  | 19/16 | 5/5 | 0/1 |
|  | Pool |  |  | 18/18 | 0/0 |
|  | Parametric |  |  |  | 13/13 |
| Interactions | Proposed | 5/5 | 0/0 | 4/4 | 0/0 |
|  | Meta |  | 7/3 | 0/0 | 0/0 |
|  | Pool |  |  | 5/5 | 0/0 |
|  | Parametric |  |  |  | 3/3 |

Note: In each cell, LUAD/LUSC.

set of identified main effects and interactions suggests that the environmental factors, i.e. age and smoking, play a non-negligible role in characterizing FEV1. Omitting these, then, can affect the identification results. Detailed estimation results using these alternative methods are provided in supplementary Appendix C.

For real data, because the main effects and interactions that influence the response variables are unknown (or, at least, partially unknown), we cannot directly evaluate and compare the identification performance of different methods. Instead, we evaluated the stability of identification and prediction. This provides some insight into the validity of the methods. Specifically, each dataset was randomly partitioned into a training set and a testing test, at a ratio of 2:1. Estimation was conducted with the training set and prediction was made with the testing set. The average prediction error over 50 independent repetitions was 0.877 (Proposed), 0.923 (Meta), 0.881 (Pool), and 1.135 (Parametric). This shows that the proposed method has higher prediction accuracy. We also evaluated the identification stability of each method using the Observed Occurrence Index (OOI). The results are given in Figure C.3 (supplementary Appendix C). The proposed method had the highest OOIs for both main effects and interactions.

**Table 5.** Analysis of NSCLC data (with survival time as the response variable) using the proposed method: estimated coefficients of main effects and interactions.

|         | LUAD   | LUSC   |                    | LUAD   | LUSC   |
|---------|--------|--------|--------------------|--------|--------|
| SBNO2   | −0.040 | −0.042 | RASGEF1A           | −0.024 | −0.097 |
| GFOD2   | −0.024 | −0.034 | WDR60              | −0.043 | −0.030 |
| PYGB    | −0.077 | −0.051 | LRWD1              | 0.034  | −0.034 |
| RIOX1   | 0.046  | 0.075  | CBR1               | 0.023  | −0.034 |
| RPRD2   | −0.032 | 0.033  | TMUB1              | −0.032 | 0.022  |
| CSPG5   | 0.065  | −0.050 | ORM2               | 0.020  | 0.021  |
| GRIN2D  | −0.029 | −0.053 | RMDN2              | −0.027 | 0.029  |
| TRIM6   | −0.041 | −0.053 | SBNO2 × RPRD2      | −0.007 | −0.006 |
| PPAN    | −0.028 | −0.069 | PYGB × RASGEF1A    | −0.009 | −0.012 |
| C7ORF57 | −0.034 | 0.025  | CSPG5 × IL6        | 0.046  | −0.057 |
| TMEM104 | −0.030 | −0.027 | TMUB1 × ORM2       | 0.009  | −0.012 |

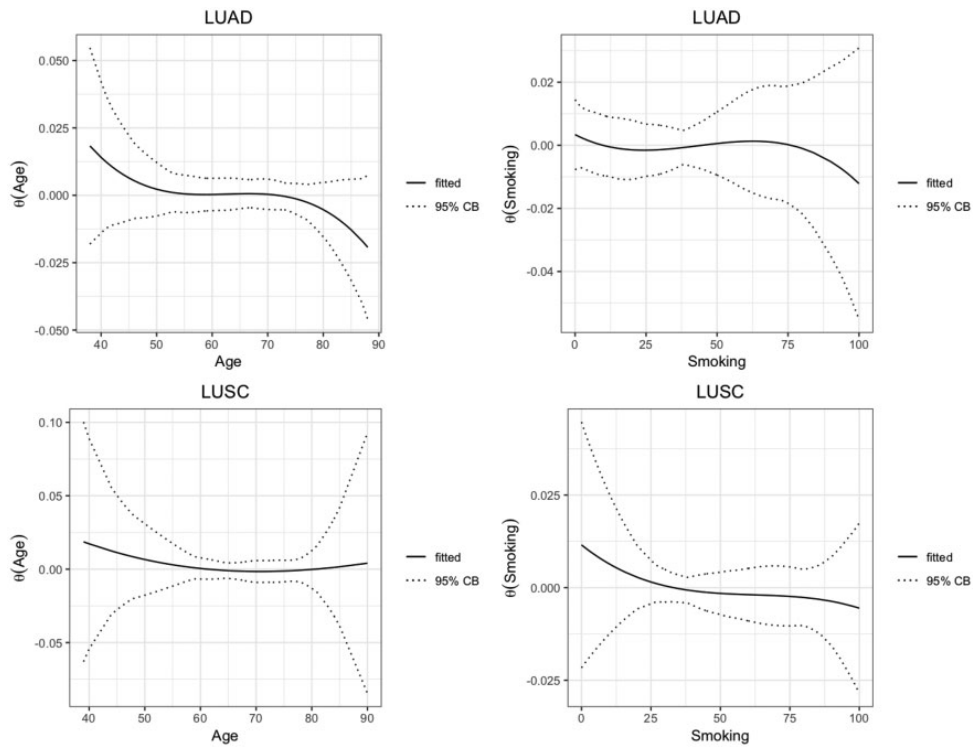

**Figure 4.** Analysis of NSCLC data (with survival time as the response variable) using the proposed method: estimated nonparametric functions of age and smoking with their 95% confidence bands.

## 4.2 Survival time as the response variable

We now investigate the influence of genetic and environmental factors on the survival time (measured in months) of NSCLC patients. Note that similar data have been analysed by Li et al.[14] Our study differed from theirs in that (1) patients of all stages were considered, and (2) the amount of smoking, measured by packs each year, was introduced as another environmental factor, in addition to age. After a series of sample selection processes (see Figure C.2 in supplementary Appendix C), 833 subjects were included in the analysis, of which the censoring rates were 64.50% (LUAD) and 56.00% (LUSC), the median age at diagnosis was 67 (range: 38–90 years), and the median amount of smoking was 40 (range: 0–240 packs/year).

We adopted the proposed method for this data under the AFT model, of which 18 main effects and four interactions were identified. Table 5 provides the estimation results of the coefficients. Once again, the

Table 6. Analysis of NSCLC data (with survival time as the response variable): number of overlapping main effects and interactions.

|              |            | Proposed | Meta  | Pool  | Parametric |
|--------------|------------|----------|-------|-------|------------|
| Main effects | Proposed   | 18/18    | 6/7   | 8/8   | 9/9        |
|              | Meta       |          | 26/24 | 5/8   | 5/2        |
|              | Pool       |          |       | 22/22 | 8/8        |
|              | Parametric |          |       |       | 13/13      |
| Interactions | Proposed   | 4/4      | 0/0   | 1/1   | 1/1        |
|              | Meta       |          | 6/2   | 0/0   | 0/0        |
|              | Pool       |          |       | 3/3   | 0/0        |
|              | Parametric |          |       |       | 3/3        |

Note: In each cell, LUAD/LUSC.

identification results were meaningful. For example, some lung cancer-related genes such as PYGB and CBR1 were identified. Over-expression of PYGB has been found in high-risk groups of lung carcinomas.[26] Moreover, CBR1 mRNA, CBR1 protein levels, and CBR1 SNPs show significant associations with the development of lung cancer.[27]

We plot the estimates of nonparametric functions in Figure 4. The survival time of LUAD patients decreased gradually as their age increased. For LUSC patients, however, the trend was slightly different: survival time decreased in patients 70 years old and younger, and increased slowly after 70 years old. This differs from observations in existing work.[14] For both LUAD and LUSC patients, their survival time decreased nonlinearly with more cigarette smoking.

The data were also analysed using the alternative methods. Table 6 shows summarized results, which again indicate that different methods lead to different results. Detailed identification and estimation results are provided in supplementary Appendix C. In terms of prediction, we used log-rank test statistics to measure the prediction accuracy, where a higher log-rank test value indicates higher prediction accuracy. Using the same resampling procedure described above, the average log-rank test statistics were 14.79 (Proposed), 12.48 (Meta), 13.69 (Pool), and 14.20 (Parametric). The results of OOIs are plotted in Figure C.4 (supplementary Appendix C). The proposed method had the highest OOIs in terms of both main effects and interactions.

### 4.3 Comparison with the penalisation method

To further gauge the performance of the proposed method, we systematically compared it to the penalisation method[14] in the NSCLC data. Because the penalisation method diverges when there are 300 genes (Figure 2), we reduced the number of genes to 100 to render the two methods comparable. Following the novel PCS framework outlined by Yu and Kumbier,[28] we evaluated the performance of the two methods in terms of prediction, computation, and stability.

To evaluate the prediction performance, we split each dataset randomly into a training and a testing set at a ratio of 3:1. This process was repeated 50 times. The average prediction errors were 0.871 (Proposed) and 0.995 (Penalisation) where the response variable was FEV1, and 4.322 (Proposed) and 4.348 (Penalisation) where the response variable was the survival time. The proposed method led to an improvement in prediction. With the same splitting approach, we also evaluated the stability of methods at effect identification. Figure C.5 (supplementary Appendix C) shows the OOI results under the linear model and AFT model. The proposed method always had a higher OOI than the penalisation method, regardless of the main effects and interactions. This suggests that the proposed method is more stable when identifying outcome-associated genetic factors. Finally, we compared the computation time of two methods. The results for a single replicate are provided in Table C.3 and Figure C.6 (supplementary Appendix C). The penalisation method needed more computation time than the proposed method, and the gap between the two methods in terms of computation time drastically increased as more genes were included in the model. For example, for $p = 150$ with the survival time as the response variable, the penalisation method required more than 15,000 s, whereas the proposed method required only 300 s. Overall, the proposed method outperformed the penalisation method with the reduced NSCLC data in terms of prediction, stability, and efficiency.

## 5 Conclusion

In this paper, we presented a semiparametric approach to jointly model multiple datasets, where genetic measurements and environmental measurements are included as parametric and nonparametric components, respectively. Gene–gene interactions were also considered. A novel TGDR method was developed to estimate the parameters of the main effects, interactions, and nonparametric functions, while meeting the strong hierarchical constraint. The same set of genetic measurements were thus identified for different datasets, but their coefficients were allowed to vary.

Simulations and analyses of the NSCLC data showed the superiority of the proposed method over alternative methods. In the simulations, the proposed method demonstrated favourable performance when identifying the main effects and interactions, and it significantly outperformed the other methods in terms of estimations and predictions. Moreover, the proposed method was robust to the simulation settings. When analysing NSCLC data under the linear model and the AFT model with right-censored data, it yielded biologically meaningful findings. The higher prediction accuracy and selection stability of the proposed approach confirmed its validity.

The proposed semiparametric model can be extended to accommodate more complex data structures, such as clustered covariates and time-varying covariates. We considered a small number of nonparametric parts and left them unselected. In future research, we will consider selecting and estimating these nonparametric parts simultaneously when there are a large number of environmental factors. Further, heterogeneous sparsity for multiple datasets can be assumed. That is, multiple models can have an overlapping but not necessarily identical set of important covariates. This can be done by modifying the update criterion for parameters in TGDR. Moreover, in this paper, nonlinear associations between environmental factors and the response variable were identified using descriptive statistics. It is worth adopting a data-driven procedure—e.g. the parametricness index (PI)[29]—to select between parametric and nonparametric models. In addition, our future work will explore the application of the proposed method to other types of cancers.

### Acknowledgements
The authors thank the editor, associate editor, and referees for their insightful comments and suggestions that have led to a significant improvement of this paper. The authors also thank Ms. Ziyuan Luo for the productive discussion.

### Declaration of conflicting interests
The author(s) declared no potential conflicts of interest with respect to the research, authorship, and/or publication of this article.

### Funding
The author(s) disclosed receipt of the following financial support for the research, authorship, and/or publication of this article: This study was supported by the National Bureau of Statistics of China (grant no. 2019LZ11), the National Natural Science Foundation of China (grant no. 71771211), and the Fund for Building World-class Universities (Disciplines) of Renmin University of China.

### ORCID iDs
Yang Li  https://orcid.org/0000-0002-6287-5094
Yifan Sun  https://orcid.org/0000-0001-7437-6244

### Supplemental material
Supplemental material for this article is available online.